\pdfoutput=1

\documentclass{article}
\usepackage{ijcai23}

\usepackage{times}
\usepackage{soul}
\usepackage{url}
\usepackage[hidelinks]{hyperref}
\usepackage[utf8]{inputenc}
\usepackage[small]{caption}
\usepackage{graphicx}
\usepackage{amsmath}
\usepackage{amsthm}
\usepackage{booktabs}
\usepackage{algorithm}
\usepackage{algorithmic}
\usepackage[switch]{lineno}

\usepackage{color}
\usepackage{amsfonts}
\newcommand{\bs}{\boldsymbol}
\newcommand{\eq}{{\rm st}}
\newcommand{\norm}{{\rm Norm}}
\newcommand{\de}{\delta}
\newcommand{\ep}{\epsilon}

\newcommand{\mc}{\mathcal}
\newtheorem{assumption}{Assumption}
\newtheorem{definition}{Definition}

\newtheorem{theorem}{Theorem}

\title{Learning in Multi-Memory Games Triggers Complex Dynamics Diverging from Nash Equilibrium\footnote{The codes that we used are available at \url{https://github.com/CyberAgentAILab/with-memory_games}}}

\author{
Yuma Fujimoto$^{1,2,3}$
\and
Kaito Ariu$^{3,4}$\And
Kenshi Abe$^3$
\affiliations
$^1$Research Center for Integrative Evolutionary Science, SOKENDAI.\\
$^2$Universal Biology Institute (UBI), the University of Tokyo.\\
$^3$AI Lab, CyberAgent, Inc.\\
$^4$KTH Royal Institute of Technology.\\
\emails
fujimoto\_yuma@soken.ac.jp,
kaito\_ariu@cyberagent.co.jp,
abe\_kenshi@cyberagent.co.jp
}

\begin{document}

\maketitle

\begin{abstract}
    Repeated games consider a situation where multiple agents are motivated by their independent rewards throughout learning. In general, the dynamics of their learning become complex. Especially when their rewards compete with each other like zero-sum games, the dynamics often do not converge to their optimum, i.e., the Nash equilibrium. To tackle such complexity, many studies have understood various learning algorithms as dynamical systems and discovered qualitative insights among the algorithms. However, such studies have yet to handle multi-memory games (where agents can memorize actions they played in the past and choose their actions based on their memories), even though memorization plays a pivotal role in artificial intelligence and interpersonal relationship. This study extends two major learning algorithms in games, i.e., replicator dynamics and gradient ascent, into multi-memory games. Then, we prove their dynamics are identical. Furthermore, theoretically and experimentally, we clarify that the learning dynamics diverge from the Nash equilibrium in multi-memory zero-sum games and reach heteroclinic cycles (sojourn longer around the boundary of the strategy space), providing a fundamental advance in learning in games.
\end{abstract}

\section{Introduction} \label{S01}
Repeated games consider that multiple agents aim to optimize their objective functions based on a normal-form game~\cite{fudenberg1991game}. It is known that in this game, the set of optimal strategies for all the agents always exists as Nash equilibria~\cite{nash1950equilibrium}. Various algorithms with which each agent achieves its optimal strategy have been proposed, such as Cross learning~\cite{cross1973stochastic}, replicator dynamics~\cite{borgers1997learning,hofbauer1998evolutionary}, gradient ascent~\cite{singh2000nash,zinkevich2003online,bowling2002multiagent,bowling2004convergence}, Q-learning~\cite{watkins1992q,kaisers2010frequency,abdallah2013addressing}, and so on. In zero-sum games where two agents have conflicts in their benefits, however, the above learning algorithms cannot converge to their equilibrium~\cite{mertikopoulos2016learning,mertikopoulos2018cycles}. Indeed, the dynamics of learning draw a loop around the equilibrium point, even though the shape of the trajectory differs more or less depending on the algorithm. Thus, solving the dynamics around the Nash equilibrium is a touchstone for discussing whether the learning works well.

Currently, several studies attempt to understand trajectories of multi-agent learning by integrating various cross-disciplinary algorithms~\cite{tuyls2005evolutionary,tuyls2006evolutionary,bloembergen2015evolutionary,barfuss2020towards}. For example, if we take an infinitesimal step size of learning, Cross learning draws the same trajectory as a replicator dynamics. The replicator dynamics can be interpreted as the weighted version of infinitesimal gradient ascent. Furthermore, Q-learning differs only in the extra term of exploration with the replicator dynamics. Another study has shown a relationship between the replicator dynamics and Q-learning by introducing a generalized regularizer which pulls the strategy back to the probabilistic simplex at the shortest distance~\cite{mertikopoulos2016learning}. Like these studies, it is important to understand the trajectory of multi-agent learning theoretically.

\begin{figure}[ht]
    \centering
    \includegraphics[width=0.8\hsize]{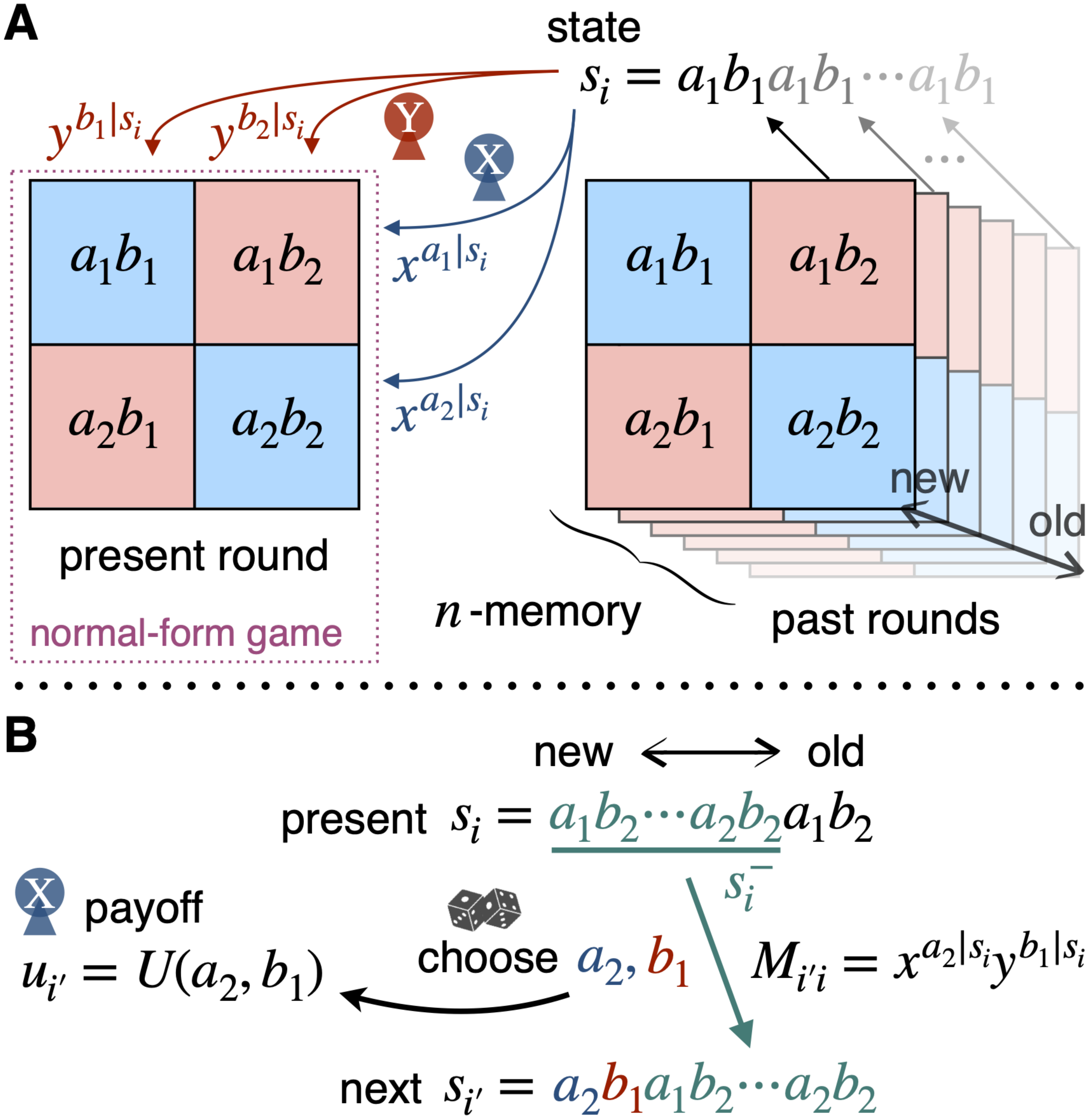}
    \caption{{\bf A}. Illustration of a multi-memory repeated game. Focusing on the area surrounded by the purple dots, a normal-form game is illustrated. Player X (resp. Y) chooses its action $a_1$ or $a_2$ in the row (resp. $b_1$ or $b_2$ in the column). Then, each of them receives its payoff depending on their actions. The panel shows the matching-pennies game, where blue (resp. red) panels show that X (resp. Y) gains a payoff of $1$ and Y (resp. X) loses it. Looking at the whole, each player memorizes their actions of the past $n$ rounds. This memorized state is described as $s_i$ given by $2n$-length bits of actions. {\bf B}. Illustration for the detailed single round of repeated games, where present state $s_i$ transitions to next state $s_{i'}$. In this transition, the oldest $2$ bits are lost, and the other bits $s_i^{-}$, colored in green, are maintained. X's and Y's choices ($a_2$ (blue) and $b_1$ (red) in this figure) are appended as the newest $2$ bits in $s_{i'}$. This transition occurs with the probability of $M_{i'i}$. Finally, X gains a payoff of $u_{i'}$ in the state transition.}
    \label{F01}
\end{figure}

Repeated games potentially include memories of agents, i.e., a possibility that agents determine their actions depending on past actions they chose (see Fig.~\ref{F01} for the illustration). Such memories can expand the choice of strategies and thus lead to the agents handling their gameplay better; for example, by reading how the other player chooses its action~\cite{fujimoto2019functional}. Indeed, agents with memories can use tit-for-tat~\cite{axelrod1981evolution} and win-stay-lose-shift~\cite{nowak1993strategy} strategies in prisoner's dilemma games, and these strategies achieve cooperation as a Nash equilibrium, explaining human behaviors. Furthermore, in the field of artificial intelligence, repeated games of agents with memory have long been of interest~\cite{sandholm1996multiagent}. Learning in memorizing past actions has also been studied in extensive-form games~\cite{zinkevich2007regret,lanctot2012no}. In economics, how a region of the Nash equilibrium is extended by multi-memory strategies is enthusiastically studied as folk theorem~\cite{fudenberg2009folk}. In practice, Q-learning is frequently implemented in multi-memory games~\cite{barfuss2019deterministic,barfuss2020reinforcement,meylahn2022limiting}. Several studies~\cite{fujimoto2019emergence,fujimoto2021exploitation} partly discuss the relation between the replicator dynamics and the gradient ascent but consider only prisoner's dilemma games. In conclusion, this relation is still unclear in games with general numbers of memories and actions. Furthermore, the convergence of dynamics in such multi-memory games has been unexplored.

This study provides a basic analysis of the multi-memory repeated game. First, we extend the two learning algorithms, i.e., replicator dynamics and gradient ascent, for multi-memory games. Then, we name them multi-memory replicator dynamics (MMRD) and gradient ascent (MMGA). As well as shown in the zero-memory games, the equivalence between MMRD and MMGA is proved in Theorems~\ref{Theorem01}-\ref{Theorem03}. Next, we tackle the convergence problem of such algorithms from both viewpoints of theory and experiment. Theorem~\ref{Theorem04} shows that under one-memory two-action zero-sum games, the Nash equilibrium is unique and essentially the same as that of zero-memory games. This theorem is nontrivial if taking into account the fact that diversification of strategies can expand the region of Nash equilibria in general games. Then, while utilizing these theorems, we see how multi-memory learning complicates the dynamics, leading to divergence from the Nash equilibrium with sensitivity to its initial condition like chaos.

\section{Preliminary} \label{S02}
\subsection{Two-Player Normal-Form Game} \label{S02-01}
Let us define two-player (of X and Y) $m(\in \mathbb{N})$-action games (see illustration of Fig.~\ref{F01}-A). Player X and Y choose their actions from $\mc{A}=\{a_1,\cdots,a_m\}$ and $\mc{B}=\{b_1,\cdots,b_m\}$ in a single round. After they finish choosing their actions $a\in\mc{A}$ and $b\in\mc{B}$, each of them gains a payoff $U(a,b)\in\mathbb{R}$ and $V(a,b)\in\mathbb{R}$, respectively.

\subsection{Two-Player Multi-Memory Repeated Game} \label{S02-02}
We further consider two-player $n(\in\mathbb{N})$-memory repeated games as an iteration of the two-player normal-form game (see illustration Fig.~\ref{F01}-A). The players are assumed to memorize their actions in the last $n$ rounds. Since each player can take $m$ actions, there are $m^{2n}$ cases for possible memorized states, described as $\mc{S}=\prod_{k=1}^{n}(\mc{A}\times\mc{B})$. Under any memorized state, player X can choose any action stochastically. Such a stochastic choice of an action is described by a parameter $x^{a|s}$, which means the probability of choosing an action $a\in\mc{A}$ under memorized state $s\in\mc{S}$. Thus, X's strategy is represented by $|\mc{S}|(=m^{2n})$-numbers of $(m-1)$-dimension simplexes, ${\bf x}\in\prod_{s\in\mc{S}}\Delta^{m-1}$, while Y's is ${\bf y}\in\prod_{s\in\mc{S}}\Delta^{m-1}$.

\subsection{Formulation as Markov Games} \label{S02-03}
In order to handle this multi-memory repeated game as a Markov game~\cite{shapley1953stochastic,littman1994markov}, we define a vector notation of memorized states;
\begin{align*}
    \bs{s}=(\underbrace{a_1b_1\cdots a_1b_1}_{\times n},\underbrace{a_1b_1\cdots a_1b_1}_{\times (n-1)}a_1b_2,\cdots, \underbrace{a_mb_m\cdots a_mb_m}_{\times n}),
\end{align*}
which orders all the elements of $\mc{S}$ as a vector. We also define a vector notation of utility function as
\begin{align*}
    \bs{u}=(&\underbrace{U(a_1,b_1),\cdots,U(a_1,b_1)}_{\times m^{2n-2}},\underbrace{U(a_1,b_2),\cdots,U(a_1,b_2)}_{\times m^{2n-2}},\\
    &\cdots,\underbrace{U(a_m,b_m),\cdots,U(a_m,b_m)}_{\times m^{2n-2}}),
\end{align*}
which orders all the last-round payoffs for $\mc{S}$ as a vector. The utility function for Y, i.e., $\bs{v}$, is defined similarly. In addition, we denote an index for these vectors as $i\in\{1,\ldots,m^{2n}\}$. $u_i$ is defined by the utility using the first 2 bits of actions in state $s_i$. For example, if $s_i=a_1b_2a_2b_1$, then $u_i=U(a_1,b_2)$.

Let $\bs{p}\in \Delta^{|\mc{S}|-1}$ be a probability distribution on $\bs{s}$ in a round. As the name Markov matrix implies, a distribution in the next round $\bs{p}'$ is given by $\bs{p}'=\bs{Mp}$, where $\bs{M}$ is a Markov transition matrix;
\begin{align}
    M_{i'i}=\left\{\begin{array}{ll}
        x^{a|s_i}y^{b|s_i} & (s_{i'}=abs_{i}^{-}) \\ 
        0 & ({\rm otherwise}) \\
    \end{array}\right.,
    \label{matrix}
\end{align}
which shows the transition probability from $i$-th state to $i'$-th one for $i,i'\in\{1,\ldots,m^{2n}\}$. Here, note that $s_{i}^{-}$ shows the state $s_i$ except for the oldest two actions. See Fig.~\ref{F01}-B illustrating an example of Markov transition.

\subsection{Nash Equilibrium} \label{S02-04}
We now analyze the Nash equilibrium in multi-memory repeated games based on the formulation of Markov games. Let us assume that every agent uses a fixed strategy ${\bf x}$ and ${\bf y}$ or learns slowly enough for the timescale of the Markov transitions. We further assume that the strategies are located within the interiors of simplexes. Under this assumption, the Markov matrix becomes ergodic, and the stationary distribution is unique, denoted as $\bs{p}^{\eq}({\bf x}, {\bf y})$. This assumption is reasonable because all the actions should be learned in the replicator dynamics, and actions that are not played cannot be learned. This stationary distribution satisfies $\bs{p}^{\eq}=\bs{M}\bs{p}^{\eq}$. We also denote each player's expected payoff in the stationary distribution as $u^{\eq}({\bf x}, {\bf y})=\bs{p}^{\eq}\cdot\bs{u}$ and $v^{\eq}({\bf x}, {\bf y})=\bs{p}^{\eq}\cdot\bs{v}$. The goal of learning in the multi-memory game is to search for the Nash equilibrium, denoted by $({\bf x}^*,{\bf y}^*)$, where their payoffs are maximized as
\begin{align}
    \left\{\begin{array}{l}
        {\bf x}^*\in\mathrm{argmax}_{{\bf x}} u^{\eq}({\bf x}, {\bf y}^*)\\
        {\bf y}^*\in\mathrm{argmax}_{{\bf y}} v^{\eq}({\bf x}^*, {\bf y})\\
    \end{array}\right..
\end{align}
Here, $u^{\eq}$ and $v^{\eq}$ are complex non-linear functions for high-dimensional variables of $({\bf x}, {\bf y})$. This Nash equilibrium is difficult to find in general.

\section{Algorithm} \label{S03}
In the following, we define multi-memory versions of two major learning algorithms, i.e., replicator dynamics and gradient ascent. Although we consider the learning of player X, that of player Y can be formulated in the same manner.

\begin{definition}[expected future payoff]
We define the expected future payoff from the distribution $\bs{p}$ as
\begin{align}
    \pi(\bs{p}, {\bf x}, {\bf y}):=&\sum_{t=0}^{\infty}\bs{M}^t(\bs{p}-\bs{p}^{\eq})\cdot\bs{u},
    \label{EFP}
\end{align}
which is the total payoff player X obtains from the present round to the future.
\label{Definition01}
\end{definition}

In this definition, the stationary payoff $\bs{p}^{\eq}\cdot\bs{u}=u^{\eq}$ is the offset term every round, and thus $\pi(\bs{p}^{\eq}, {\bf x}, {\bf y})=0$.

\begin{definition}[normalization]
We define the normalization function $\norm:\prod_{s\in\mc{S}}\mathbb{R}_+^m\mapsto\prod_{s\in\mc{S}}\mathrm{int}(\Delta^{m-1})$ as
\begin{align}
    \norm({\bf x})=\left\{\frac{x^{a|s}}{\sum_{a'}x^{a'|s}}\right\}_{a,s},
\end{align}
\end{definition}

In this definition, $\norm(\bf{x})$ satisfies the condition of probability variables for all $s$.

Based on these definitions, we formulate discretized MMRD and MMGA as Algorithm~\ref{D-MMRD} and~\ref{D-MMGA}.

\begin{algorithm}[H]
    \caption{Discretized MMRD}
    \label{D-MMRD}
    \textbf{Input}: $\eta$
    \begin{algorithmic}[1]
        \FOR{$t=0,1,2,\cdots$}
        \label{A1-L01}
        \STATE X chooses $a$ with probability $x^{a|s_i}$
        \label{A1-L02}
        \STATE (Y chooses $b$ with probability $y^{b|s_i}$)
        \label{A1-L03}
        \STATE $s_{i'}\gets abs_i^{-}$
        \label{A1-L04}
        \STATE $x^{a|s_i}\gets x^{a|s_i}+\eta \pi(\bs{e}_{i'},{\bf x},{\bf y})$
        \label{A1-L05}
        \STATE ${\bf x}\gets \norm({\bf x})$
        \label{A1-L06}
        \STATE $s_i\gets s_{i'}$
        \label{A1-L07}
        \ENDFOR
        \label{A1-L08}
    \end{algorithmic}
\end{algorithm}
Algorithm~\ref{D-MMRD} (Discretized MMRD) takes its learning rate $\eta$ as an input. In each time step, the players choose their actions following their strategies (lines \ref{A1-L02} and \ref{A1-L03}), while the state is updated by their chosen actions (lines \ref{A1-L04} and \ref{A1-L07}). In line \ref{A1-L05}, each player reinforces its strategy by how much payoff it receives in the future from state $s_{i'}$. Here, note that $\bs{e}_{i'}$ indicates the unit vector for the $i'$-th element, describing that state $s_{i'}$ occurs.

\begin{algorithm}[H]
    \caption{Discretized MMGA}
    \label{D-MMGA}
    \textbf{Input}: $\eta$, $\gamma$
    \begin{algorithmic}[1]
        \FOR{$t=0,1,2,\cdots$}
        \label{A2-L01}
        \FOR{$a\in\mc{A}$, $s\in\mc{S}$}
        \label{A2-L02}
        \STATE ${\bf x}'\gets \norm({\bf x}+\gamma{\bf e}^{a|s})$
        \label{A2-L03}
        \STATE $\displaystyle \Delta^{a|s}\gets\frac{u^{\eq}({\bf x}',{\bf y})-u^{\eq}({\bf x},{\bf y})}{\gamma}$
        \label{A2-L04}
        \ENDFOR
        \label{A2-L05}
        \FOR{$a\in\mc{A}$, $s\in\mc{S}$}
        \label{A2-L06}
        \STATE $x^{a|s}\gets x^{a|s}(1+\eta\Delta^{a|s})$
        \label{A2-L07}
        \ENDFOR
        \label{A2-L08}
        \STATE ${\bf x}\gets\norm({\bf x})$
        \label{A2-L09}
        \ENDFOR
        \label{A2-L10}
    \end{algorithmic}
\end{algorithm}
Algorithm~\ref{D-MMGA} (Discretized MMGA) takes not only its learning rate $\eta$ but a small value $\gamma$ in measuring an approximate gradient as inputs. In each time step, each player measures the gradients of its payoff for each variable of its strategy (lines \ref{A2-L02}-\ref{A2-L05}). Here, ${\bf e}^{a|s}$ is an abused notation of unit vector for the element of action $a$ for state $s$. Then, the player updates its strategy by the gradients (lines \ref{A2-L06}-\ref{A2-L09}). Here, note that the strategy update is weighted by the probability $x^{a|s}$ (line \ref{A2-L07}) in order to correspond to Algorithm~\ref{D-MMRD}. Here, lines \ref{A2-L03}-\ref{A2-L04} can be parallelized for all $a$ and $s$, and line \ref{A2-L07} as well.

\section{Theoretical Analysis} \label{S04}
\subsection{Continuous-Time Equivalence of Algorithms} \label{S04-01}
The following theorems provide a unified understanding of different algorithms. Theorem~\ref{Theorem01} and \ref{Theorem02} are concerned with continualization of the two discrete algorithms. Surprisingly, Theorem~\ref{Theorem03} proves the correspondence between these different continualized algorithms by Theorem~\ref{Theorem01} and \ref{Theorem02}.

\begin{theorem}[Coutinualized MMRD]
Let $\bs{p}^{a|s}$ be the expected distribution when X chooses $a$ under state $s$;
\begin{align}
    p^{a|s}_{i'}:=\left\{\begin{array}{ll}
        y^{b|s} & (s_{i'}=abs^{-}) \\
        0 & (\mathrm{otherwise}) \\
    \end{array}\right..
\end{align}
In the limit of $\eta\to 0$, Algorithm~\ref{D-MMRD} is continualized as dynamics
\begin{align}
    \dot{x}^{a|s_i}({\bf x},{\bf y})&=p_{i}^{\eq}x^{a|s_i}\left(\pi(\bs{p}^{a|s_i}, {\bf x}, {\bf y})-\bar{\pi}^{s_i}({\bf x}, {\bf y})\right),
    \label{C-MMRD_1}\\
    \bar{\pi}^{s_i}({\bf x}, {\bf y})&=\sum_{a}x^{a|s_i}\pi(\bs{p}^{a|s_i}, {\bf x}, {\bf y}),
    \label{C-MMRD_2}
\end{align}
for all $a\in\mc{A}$ and $s_i\in\mc{S}$. Here, $\bar{\pi}^{s_i}$ is the expected payoff under state $s_i$.
\label{Theorem01}
\end{theorem}

\begin{theorem}[Continualized MMGA]
In the limit of $\gamma\to 0$ and $\eta\to 0$, Algorithm~\ref{D-MMGA} is continualized as dynamics
\begin{align}
    \dot{x}^{a|s}({\bf x}, {\bf y})=x^{a|s}\frac{\partial}{\partial x^{a|s}}u^{\eq}(\norm({\bf x}),{\bf y}),
    \label{C-MMGA}
\end{align}
for all $a\in\mc{A}$ and $s\in\mc{S}$.
\label{Theorem02}
\end{theorem}

See Technical Appendix~\ref{AS01-01} and \ref{AS01-02} for the proof of Theorems~\ref{Theorem01} and~\ref{Theorem02}.

\begin{theorem}[Equivalence between the algorithms]
The dynamics Eqs.~\eqref{C-MMRD_1} and \eqref{C-MMGA} are equivalent.
\label{Theorem03}
\end{theorem}

\noindent{\it Proof Sketch.} Let ${\bf x}'$ be the strategy given by $x^{a|s}\gets x^{a|s}+\gamma$ in ${\bf x}$ for $a\in\mc{A}$ and $s\in\mc{S}$. Then, we consider the changes of the Markov transition matrix $\mathrm{d}\bs{M}:=\bs{M}(\norm({\bf x}'),{\bf y})-\bs{M}({\bf x},{\bf y})$ and the stationary distribution $\mathrm{d}\bs{p}^{\eq}:=\bs{p}^{\eq}(\norm({\bf x}'),{\bf y})-\bs{p}^{\eq}({\bf x},{\bf y})$. By considering this changes in the stationary condition $\bs{p}^{\eq}=\bs{M}\bs{p}^{\eq}$, we get $\mathrm{d}\bs{p}^{\eq}=(\bs{E}-\bs{M})^{-1}\mathrm{d}\bs{M}\bs{p}^{\eq}$ in $O(\gamma)$. The right-hand (resp. left-hand) side of this equation corresponds to the continualized MMRD (resp. MMGA). \qed

For games with a general number of actions, the study \cite{zinkevich2003online} has proposed a gradient ascent algorithm in relation to replicator dynamics. In light of this study, Theorem~\ref{Theorem03} extends the relation to the multi-memory games. This extension is neither simple nor trivial. The relation between replicator dynamics and gradient ascent has been proved by directly calculating $u^{\eq}=\bs{p}^{\eq}\cdot\bs{u}$~\cite{bloembergen2015evolutionary}. In multi-memory games, however, $u^{\eq}=\bs{p}^{\eq}\cdot\bs{u}$ is too hard to calculate. Thus, as seen in the proof sketch, we proved the relation by considering a slight change in the stationary condition $\bs{p}^{\eq}=\bs{M}\bs{p}^{\eq}$, technically avoiding such a hard direct calculation.

\subsection{Learning Dynamics Near Nash Equilibrium} \label{S04-02}
Below, let us discuss the learning dynamics in multi-memory games, especially divergence from the Nash equilibrium in zero-sum payoff matrices. In order to obtain a phenomenological insight into the learning dynamics simply, we assume one-memory two-action zero-sum games in Assumption~\ref{Assumption01}.

\begin{assumption}[One-memory two-action zero-sum game]
We assume a two-action (i.e., $\mc{A}=\{a_1,a_2\}$ and $\mc{B}=\{b_1,b_2\}$), one-memory (i.e., $\bs{s}=(a_1b_1,a_1b_2,a_2b_1,a_2b_2)$), and zero-sum game (i.e., $\bs{v}=-\bs{u}$). In particular, we discuss zero-sum games where both $u_1$ and $u_4$ are smaller or larger than both $u_2$ and $u_3$.
\label{Assumption01}
\end{assumption}

Under Assumption~\ref{Assumption01}, we exclude uninteresting zero-sum payoff matrices that the Nash equilibrium exists as a set of pure strategies because the learning dynamics trivially converge to such pure strategies. The condition that both $u_1$ and $u_4$ are smaller or larger than both $u_2$ and $u_3$ is necessary and sufficient for the existence of no dominant pure strategy.

In the rest of this paper, we use a vector notation for strategies of X and Y; $\bs{x}:=\{x_i\}_{i=1,\ldots, 4}$ and $\bs{y}:=\{y_i\}_{i=1,\ldots, 4}$ as $x_{i}:=x^{a_1|s_i}$ and $y_{i}:=y^{b_1|s_i}$. Indeed, $x^{a_2|s_i}=1-x_{i}$ and $y^{b_2|s_i}=1-y_{i}$ hold.

\begin{theorem}[Uniqueness of the Nash equilibrium]
Under Assumption~\ref{Assumption01}, the unique Nash equilibrium of this game is $(x_{i},y_{i})=(x^*,y^*)$ for all $i$ as
\begin{align}
    x^*=\frac{-u_3+u_4}{u_1-u_2-u_3+u_4},\ y^*=\frac{-u_2+u_4}{u_1-u_2-u_3+u_4}.
\end{align}
\label{Theorem04}
\end{theorem}

\noindent{\it Proof Sketch.} Let us prove that X's strategy in the Nash equilibrium is uniquely $\bs{x}=x^*\bs{1}$. First, we define $u^*$ and $v^*$ as X's and Y's payoffs in the Nash equilibrium in the zero-memory game. If $\bs{x}=x^*\bs{1}$, X's expected payoff is $u^{\eq}=u^*$, regardless of Y's strategy $\bs{y}$. Second, we consider that X uses another strategy $\bs{x}\neq x^*\bs{1}$. Then, there is Y's strategy such that $v^{\eq}>v^*\Leftrightarrow u^{\eq}<u^*$. Thus, X's minimax strategy is uniquely $\bs{x}=x^*\bs{1}$, completing the proof. \qed
\begin{figure}[ht]
    \centering
    \includegraphics[width=0.9\hsize]{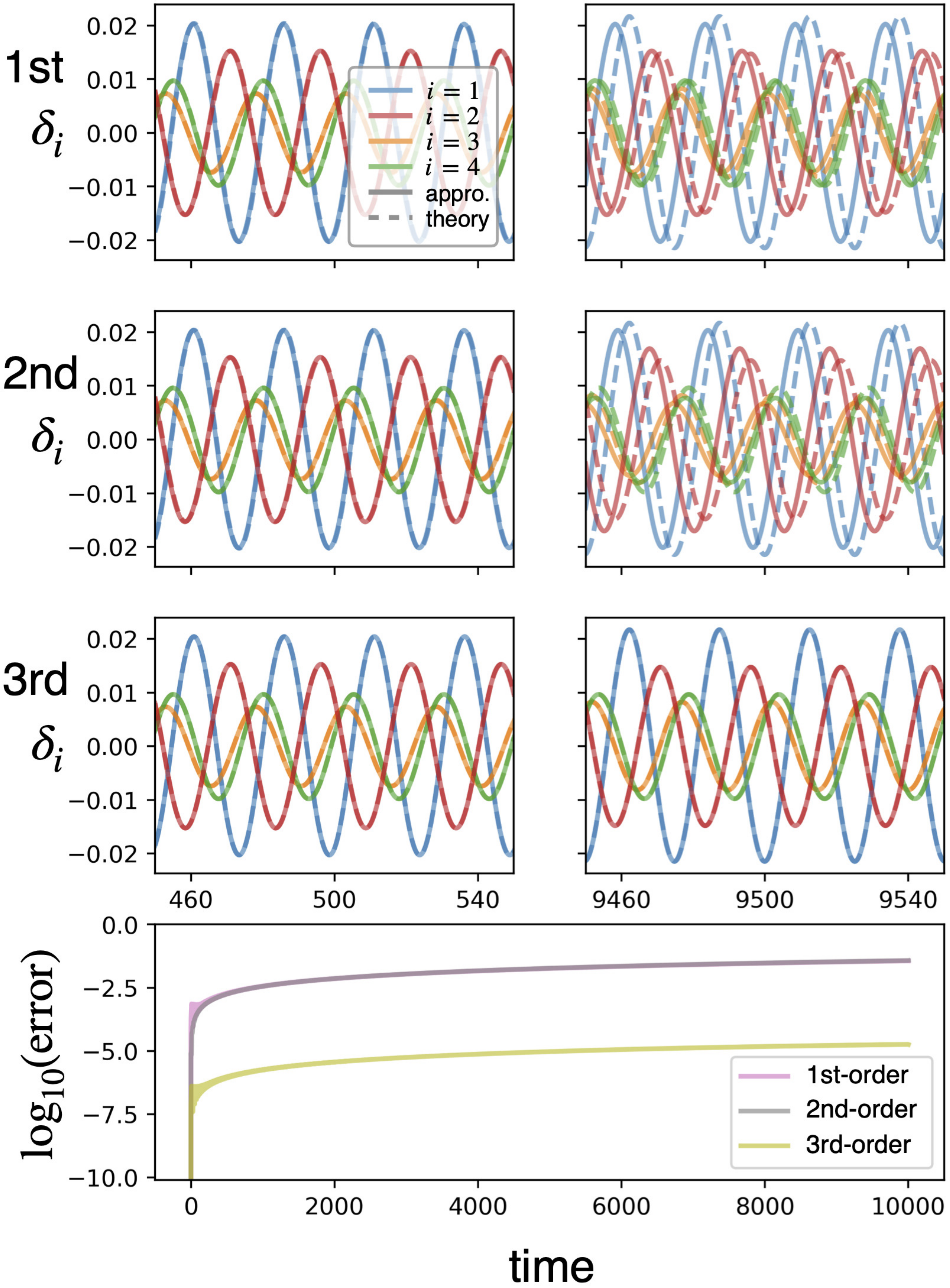}
    \caption{Multi-memory learning dynamics near the Nash equilibrium in the matching-pennies game. In the upper six panels, colored lines indicate the time series of $\delta_i$ (X's strategy). The solid (resp. broken) lines are approximated (resp. experimental) trajectories of learning dynamics. From the top, the trajectories are predicted by approximations up to the first, second, and third orders. The bottom panel shows the errors between the approximated and experimental trajectories.}
    \label{F02}
\end{figure}

\begin{figure*}[ht]
    \centering
    \includegraphics[width=0.75\hsize]{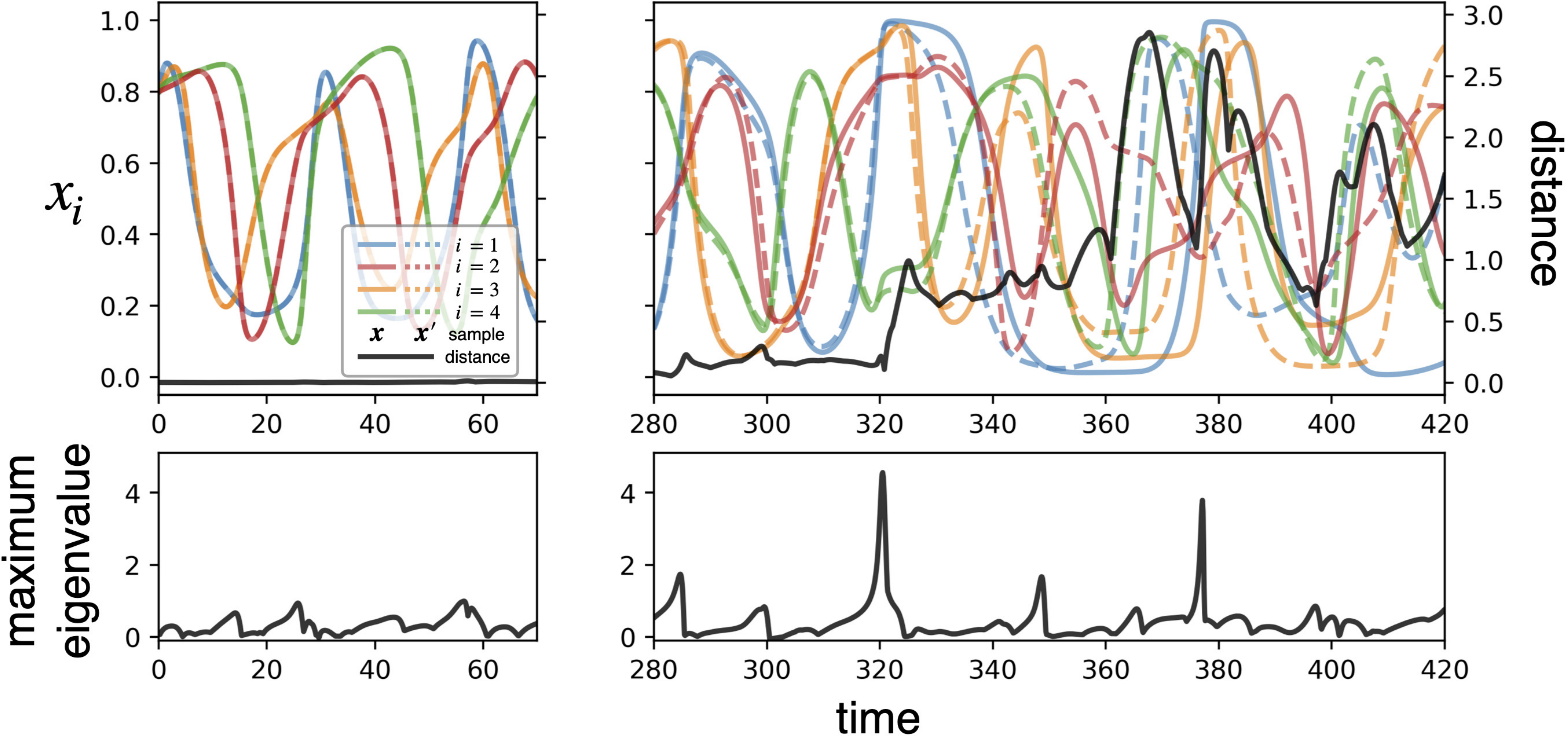}
    \caption{Initial state sensitivity in learning dynamics in multi-memory games. In the top panels, colored lines are time series of $x_i$ (X's strategy). The black line is the distance between the solid (sample of $\bs{x}$) and broken ($\bs{x}'$) lines. In the bottom panels, the black lines indicate the maximum eigenvalue in the learning dynamics of the solid line.}
    \label{F03}
\end{figure*}

Regarding Theorem~\ref{Theorem04}, X (Y) chooses each action in the same probability independent of the last state. Here, they do not utilize their memory. Thus, note that in this sense, the Nash equilibrium is the same as that in the zero-memory version of the game. This theorem means that in zero-sum games, the region of the Nash equilibrium does not expand even if players have memories. Taking into account that having multiple memories expands the region of Nash equilibria, such as a cooperative equilibrium in prisoner's dilemma games~\cite{axelrod1981evolution}, this theorem is nontrivial.

In order to discuss whether our algorithms converge to this unique Nash equilibrium under Assumption~\ref{Assumption01}, we consider the neighbor of the Nash equilibrium and define sufficient small deviation from the Nash equilibrium, i.e., $\bs{\de}:=\bs{x}-x^*\bs{1}$ and $\bs{\ep}:=\bs{y}-y^*\bs{1}$. Here, we assume that these deviations have the same scale $O(\de):=O(\de_i)=O(\ep_i)$ for all $i$. Then, defining that the superscript $(k)$ shows $O(\de^{k})$ terms, the dynamics are approximated by $\dot{\bs{x}}\simeq\dot{\bs{x}}^{(1)}+\dot{\bs{x}}^{(2)}$ and $\dot{\bs{y}}\simeq\dot{\bs{y}}^{(1)}+\dot{\bs{y}}^{(2)}$;
\begin{align}
    \dot{\bs{x}}^{(1)}&=+x^*(1-x^*)(\bs{u}\cdot\bs{1}_{\sf{z}})\bs{p}^*\circ\bs{\ep},
    \label{dotx(1)}\\
    \dot{\bs{y}}^{(1)}&=-y^*(1-y^*)(\bs{u}\cdot\bs{1}_{\sf{z}})\bs{p}^*\circ\bs{\de},
    \label{doty(1)}\\
    \dot{\bs{x}}^{(2)}&=-(x^*-\tilde{x}^*)(\bs{u}\cdot\bs{1}_{\sf{z}})\bs{\de}\circ\bs{\ep}\circ\bs{p}^*
    \nonumber\\
    &\hspace{0.4cm}+x^*\tilde{x}^*(\bs{u}\cdot\bs{1}_{\sf{z}})\{(\bs{\de}\cdot\bs{p}^*)\bs{\ep}\circ\bs{y}^*\circ\bs{1}_{\sf{x}}
    \nonumber\\
    &\hspace{0.5cm}+(\bs{\ep}\cdot\bs{p}^*)\bs{\ep}\circ\bs{x}^*\circ\bs{1}_{\sf{y}}+(\bs{\de}\circ\bs{\ep}\circ\bs{y}^*\cdot\bs{1}_{\sf{x}})\bs{p}^*\},
    \label{dotx(2)}\\
    \dot{\bs{y}}^{(2)}&=+(y^*-\tilde{y}^*)(\bs{u}\cdot\bs{1}_{\sf{z}})\bs{\de}\circ\bs{\ep}\circ\bs{p}^*
    \nonumber\\
    &\hspace{0.4cm}-y^*\tilde{y}^*(\bs{u}\cdot\bs{1}_{\sf{z}})\{(\bs{\de}\cdot\bs{p}^*)\bs{\de}\circ\bs{y}^*\circ\bs{1}_{\sf{x}}
    \nonumber\\
    &\hspace{0.5cm}+(\bs{\ep}\cdot\bs{p}^*)\bs{\de}\circ\bs{x}^*\circ\bs{1}_{\sf{y}}+(\bs{\de}\circ\bs{\ep}\circ\bs{x}^*\cdot\bs{1}_{\sf{y}})\bs{p}^*\},
    \label{doty(2)}
\end{align}
with $\bs{x}^*:=(x^*,x^*,\tilde{x}^*,\tilde{x}^*)$, $\bs{y}^*:=(y^*,\tilde{y}^*,y^*,\tilde{y}^*)$, $\bs{p}^*:=\bs{x}^*\circ\bs{y}^*$, $\bs{1}_{\sf{x}}:=(+1,+1,-1,-1)$, $\bs{1}_{\sf{y}}:=(+1,-1,+1,-1)$, and $\bs{1}_{\sf{z}}:=\bs{1}_{\sf{x}}\circ\bs{1}_{\sf{y}}$. Eqs.~\eqref{dotx(1)}-\eqref{doty(2)} are derived by considering small changes in the stationary condition $\bs{p}^{\eq}=\bs{M}\bs{p}^{\eq}$ for deviations of $\bs{\de}$ and $\bs{\ep}$ (see Technical Appendix~\ref{AS02-01} and \ref{AS02-02} for the detailed calculation). By that, we can avoid a direct calculation of $\bs{p}^{\eq}$, which is hard to be obtained.

\section{Experimental Findings} \label{S05}
\subsection{Simulation and Low-Order Approximation} \label{S05-01}
From the obtained dynamics, i.e., Eqs.~\eqref{dotx(1)}-\eqref{doty(2)}, we interpret the learning dynamics in detail. In the first-order dynamics, multi-memory learning is no more than a simple extension of the zero-memory one. Indeed, the zero-memory learning draws an elliptical orbit given by Hamiltonian as the conserved quantity~\cite{hofbauer1996evolutionary,mertikopoulos2018cycles}. Eqs.~\eqref{dotx(1)} and \eqref{doty(1)} mean that the multi-memory dynamics also draw similar elliptical orbits for each pair of $x_i$ and $y_i$. In other words, the dynamics are given by a linear flow on a four-dimensional torus. Because no interaction occurs between the pair of $i$ and $i'$ such that $i\neq i'$, the dynamics of the multi-memory learning for each state are qualitatively the same as learning without memories. Fig.~\ref{F02} shows the time series of the multi-memory learning dynamics near the Nash equilibrium in an example of a two-action zero-sum game, the matching-pennies game ($u_1=u_4=1$, $u_2=u_3=-1$). The experimental trajectories are generated by the Runge-Kutta fourth-order method of Eq.~\eqref{C-MMGA} (see Technical Appendix~\ref{AS02-03} for details), while the approximated trajectories are by the Runge-Kutta fourth-order method for the first- (Eqs.~\eqref{dotx(1)} and \eqref{doty(1)}), the second- (Eqs.~\eqref{dotx(2)} and \eqref{doty(2)}), and the third-order approximations (in Technical Appendix~\ref{AS02-02}). The step-size is $10^{-2}$ in common. The top-left panel in the figure shows that the dynamics roughly draw a circular orbit for each state and are well approximated by the first-order dynamics of Eqs.~\eqref{dotx(1)} and \eqref{doty(1)}. However, the top-right panel, where a sufficiently long time has passed, shows that the dynamics deviate from the circular orbits (see Technical Appendix~\ref{AS03} in detail).

\begin{figure*}[ht]
    \centering
    \includegraphics[width=1.0\hsize]{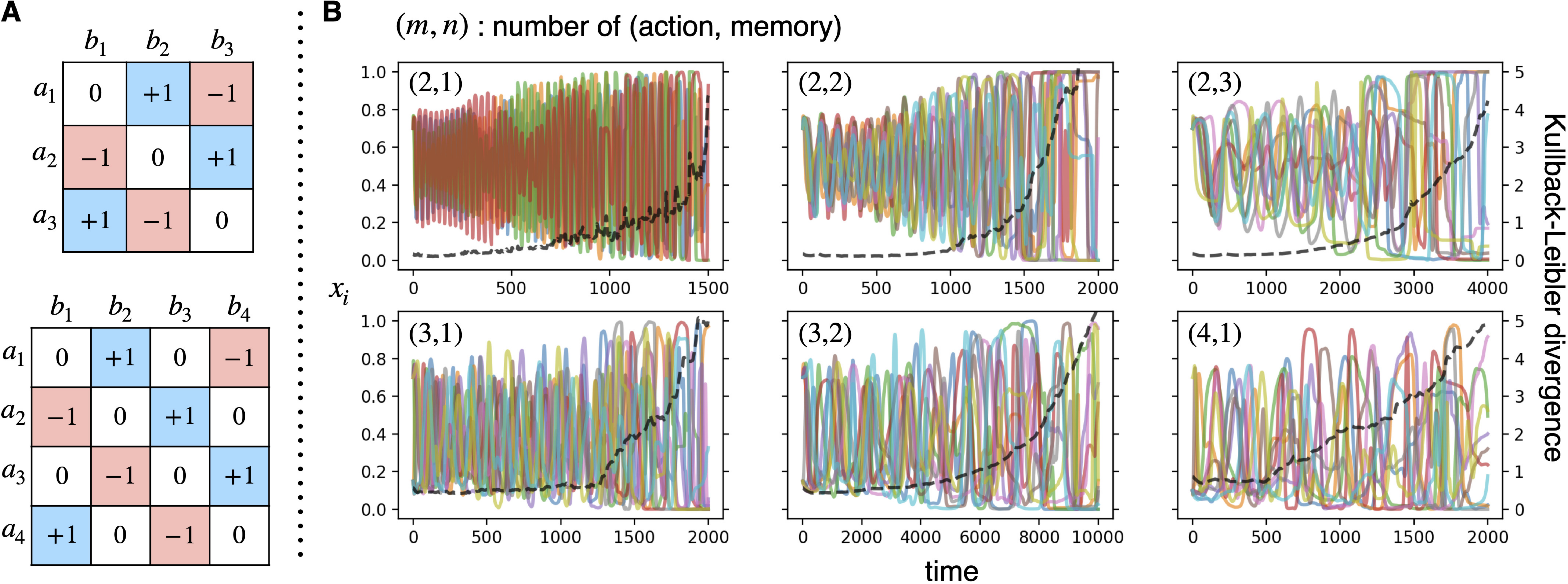}
    \caption{{\bf A}. Payoff matrices of three-action (rock-paper-scissors) and four-action (extended rock-paper-scissors) games. {\bf B}. In each panel, colored lines indicate time series of $x^{a|s}$ for random $a\in\mc{A}$ and $s\in\mc{S}$. The black broken line indicates the Kullback-Leibler divergence averaged over all the states $s\in\mc{S}$, intuitively meaning a distance from the Nash equilibrium.}
    \label{F04}
\end{figure*}

Such deviation from the circular orbits is given by higher-order dynamics than Eqs.~\eqref{dotx(1)} and \eqref{doty(1)}. In the second-order dynamics given by Eqs.~\eqref{dotx(2)} and \eqref{doty(2)}, the multi-memory learning is qualitatively different from the zero-memory one. Indeed, Eqs.~\eqref{dotx(2)} and \eqref{doty(2)} obviously mean that interactions occur between the pair of $i$ and $i'$ such that $i\neq i'$. Thereby, the dynamics of multi-memory learning become much more complex than that of zero-memory learning. In practice, no Hamiltonian function, denoted by $H^{(2)}$, exists in the second-order dynamics, as different from the first-order one. One can check this by calculating $\partial\dot{x}_{i}^{(2)}/\partial\ep_{i'}+\partial \dot{y}_{i'}^{(2)}/\partial\de_{i}\neq 0$ for $i$ and $i'\neq i$, if assuming that Hamiltonian should satisfy $\dot{\bs{x}}^{(2)}=+\partial H^{(2)}/\partial\bs{\ep}$ and $\dot{\bs{y}}^{(2)}=-\partial H^{(2)}/\partial\bs{\de}$. Thus, the multi-memory dynamics might not have any conserved quantities and not draw any closed trajectory. Indeed, the right panels in Fig.~\ref{F02} show that the dynamics tend to diverge from the Nash equilibrium. 
This divergence from the Nash equilibrium is surprising because zero-memory learning in zero-sum games always has a closed trajectory and keeps the Kullback-Leibler divergence from the Nash equilibrium constant~\cite{piliouras2014persistent,mertikopoulos2018cycles}. 
Here, note that we need the third-order dynamics to fit the experimental dynamics well, as seen by comparing the middle-right and lower-right panels in Fig.~\ref{F02}. The error between the experiment ($\bs{\delta}$ and $\bs{\epsilon}$) and approximation ($\bs{\delta}'$ and $\bs{\epsilon}'$) is evaluated by
\begin{align}
    \mathrm{error}:=\frac{1}{4}\sum_{i=1}^{4}\sqrt{|\delta_i-\delta'_i|^2+|\epsilon_i-\epsilon'_i|^2}.
\end{align}

\subsection{Chaos-Like and Heteroclinic Dynamics} \label{S05-02}
Interestingly, learning dynamics in multi-memory games are complex. Fig.~\ref{F03} shows two learning dynamics between which there is a slight difference in their initial strategies ($\bs{x}=\bs{y}=0.8\times\bs{1}$ in the solid line, but in the broken line ($\bs{x}'$ and $\bs{y}'$), $x'_1=0.801$ and others are the same as the solid line). We use Algorithm~\ref{D-MMGA} with $\eta=10^{-3}$ and $\gamma=10^{-6}$. These dynamics are similar in the beginning ($0\le t\le 320$). However, the difference between these dynamics is gradually amplified ($320\le t\le 360$), leading to the crucial difference eventually ($360\le t\le 420$). We here introduce the distance between $\bs{x}'$ and $\bs{x}$ as
\begin{align}
    D(\bs{x}',\bs{x}):=\frac{1}{4}\sum_{i=1}^{4}|L(x'_i)-L(x_i)|,
\end{align}
with $L(x):=\log x-\log(1-x)$; $L(x)$ is the measure taking into account the weight in replicator dynamics. Furthermore, in order to analyze how the difference is amplified, Fig.~\ref{F03} also shows the maximum eigenvalue in learning dynamics. We can see that the larger the maximum eigenvalue is, the more the difference between the two trajectories is amplified. We observe that such an amplification typically occurs when strategies are close to the boundary of the simplex. In conclusion, the learning dynamics provide chaos-like sensitivity to the initial condition.

\subsection{Divergence in General Memories and Actions} \label{S05-03}
Although we have focused on the one-memory two-action zero-sum games so far, numerical simulations demonstrate that similar phenomena are seen in games of other numbers of memories and actions. Fig.~\ref{F04} shows the trajectories of learning dynamics in various multi-memory and multi-action games, where we use Algorithm~\ref{D-MMGA} with $\eta=10^{-2}$ and $\gamma=10^{-6}$. Note that we consider zero-sum games in all the panels (see Fig.~\ref{F04}-A for the payoff matrices). In Fig.~\ref{F04}-B, each panel shows that strategy variables $x^{a|s}$ roughly diverge from the Nash equilibrium and sojourn longer at the edges of the simplex, i.e., $x^{a|s}=0$ or $1$. Furthermore, Kullback-Leibler divergence from the Nash equilibrium averaged over the whole states, i.e.,
\begin{align}
    D_{\rm KL}({\bf x}^*\|{\bf x}):=\frac{1}{|\mc{S}|}\sum_{s\in\mc{S}}\sum_{a\in\mc{A}}x^{*a|s}\log\frac{x^{*a|s}}{x^{a|s}}.
\end{align}
also increases with time in each panel of the figure. Thus, we confirm that learning reaches heteroclinic cycles under various (action, memory) pairs.

\section{Conclusion} \label{S06}
This study contributes to an understanding of a cutting-edge model of learning in games in Sections 3 and 4. In practice, several famous algorithms, i.e., replicator dynamics and gradient ascent, were newly extended to multi-memory games (Algorithms~\ref{D-MMRD} and \ref{D-MMGA}). We proved the correspondence between these algorithms (Theorems~\ref{Theorem01}-\ref{Theorem03}) in general. Under the assumptions of one-memory two-action zero-sum games, we further proved the uniqueness of the Nash equilibrium in two-action zero-sum games (Theorem~\ref{Theorem04}). As a background, even if agents do not have their memories, multi-agent learning dynamics are generally complicated. Thus, many theoretical approaches usually have been taken to grasp such complicated dynamics. Learning dynamics in multi-memory games are much more complicated and the dimension of strategy space of an agent explodes as $(m-1)m^{2n}$ with memory number $n$ and action number $m$. Despite these challenges, our theorems succeeded in capturing chaos-like and diverging behaviors of the dynamics. Potential future studies may focus on considering how to avoid the curse of dimension in the strategy space and proving whether the Nash equilibrium is unique in general numbers of action and memory.

This study also experimentally discovered a novel and non-trivial phenomenon that simple learning algorithms such as replicator dynamics and gradient ascent asymptotically reaches a heteroclinic cycle in multi-memory zero-sum games. In other words, the players choose actions in highly skewed proportions throughout learning. Such a phenomenon is specific to multi-memory games: Perhaps this is because the gameplay becomes extreme in learning between those who can use equally sophisticated (i.e., multi-memory) strategies. We also found a novel problem that the Nash equilibrium is difficult to reach in multi-memory zero-sum games. Here, note that convergence to the Nash equilibrium, either as a last-iterate~\cite{daskalakis2018training,daskalakis2018last,mertikopoulos2018optimistic,golowich2020tight,wei2021linear,lei2021last,abe2022mutation} or as an average of trajectories~\cite{banerjee2005efficient,zinkevich2007regret,daskalakis2011near}, is a frequently discussed topic. In general, heteroclinic cycles fail to converge even on average. What algorithm can converge to the Nash equilibrium in multi-memory zero-sum games would be interesting future work.

\section*{Acknowledgments}
We thank Tetsuro Morimura and Kunihiko Kaneko for fruitful discussions. Y.F. acknowledges the support by JSPS KAKENHI Grant No. JP21J01393.



\newpage

\onecolumn
\appendix

\newcommand{\rd}{\mathrm{d}}

\renewcommand{\theequation}{A\arabic{equation}}
\setcounter{equation}{0}
\renewcommand{\figurename}{FIG. A}
\setcounter{figure}{0}
\renewcommand{\thealgorithm}{A\arabic{algorithm}}
\setcounter{algorithm}{0}

\begin{center}
{\LARGE {\bf Appendix}}
\end{center}

\section{Proofs} \label{AS01}
\subsection{Proof of Theorem 1} \label{AS01-01}
First, line~\ref{A1-L06} in Algorithm~\ref{D-MMRD} is equal to
\begin{align}
    x^{a'|s_i}\gets\left\{\begin{array}{ll}
        x^{a'|s_i}+(1-x^{a'|s_i})\eta\pi(\bs{e}_{i'},{\bf x},{\bf y})+O(\gamma^2) & (a'=a) \\
        x^{a'|s_i}-x^{a'|s_i}\eta\pi(\bs{e}_{i'},{\bf x},{\bf y})+O(\gamma^2) & (a'\neq a) \\
    \end{array}\right.,
\end{align}
from Definition~\ref{Definition01}.

In the stationary state of the repeated games, state $s_i$ occurs with the probability of $p^{\eq}_{i}$. Then, player X (resp. Y) chooses action $a$ (resp. $b$) with the probability of $x^{a|s_i}$ and $y^{b|s_i}$. If we take the limit $\eta\to 0$ for updating $1/\eta$ times, Algorithm~\ref{D-MMRD} is continualized as dynamics
\begin{align}
    \dot{x}^{a|s_i}&=p^{\eq}_{i}\sum_{b}y^{b|s_i}\Biggl(x^{a|s_i}(1-x^{a|s_i})\pi(\bs{e}_{i'(a,b)},{\bf x},{\bf y})+\sum_{a'\neq a}x^{a'|s_i}(-x^{a|s_i})\pi(\bs{e}_{i'(a',b)},{\bf x},{\bf y})\Biggr)\\
    &=p^{\eq}_{i}x^{a|s}\Biggl\{\pi(\underbrace{\sum_{b}y^{b|s_i}\bs{e}_{i'(a,b)}}_{=\bs{p}^{a|s}},{\bf x},{\bf y})-\sum_{a'}x^{a'|s_i}\pi(\underbrace{\sum_{b}y^{b|s_i}\bs{e}_{i'(a',b)}}_{=\bs{p}^{a'|s_i}},{\bf x},{\bf y})\Biggr\}\\
    &=p^{\eq}_{i}x^{a|s_i}\Biggl(\pi(\bs{p}^{a|s_i},{\bf x},{\bf y})-\underbrace{\sum_{a'}x^{a'|s_i}\pi(\bs{p}^{a'|s_i},{\bf x},{\bf y})}_{=\bar{\pi}^{s_i}({\bf x},{\bf y})}\Biggr).
    \label{proof1_4}
\end{align}
Here, $i'(a,b)$ indicates the next state index $i'$ such that $s_{i'}=abs_{i}^{-}$. Eq.~\eqref{proof1_4} corresponds to Eqs~\eqref{C-MMRD_1} and~\eqref{C-MMRD_2} in the main manuscript. \qed

\subsection{Proof of Theorem 2} \label{AS01-02}
Taking the limit $\gamma\to 0$, we obtain
\begin{align}
    \Delta^{a|s}=\frac{\partial u^{\eq}(\norm({\bf x}),{\bf y})}{\partial x^{a|s}}.
\end{align}
Then, if we take the limit $\eta\to 0$ for updating $1/\eta$ times, Algorithm~2 is continualized as dynamics
\begin{align}
    \dot{x}^{a|s}({\bf x}, {\bf y})=x^{a|s}\frac{\partial}{\partial x^{a|s}}u^{\eq}(\norm({\bf x}),{\bf y}).
    \label{proof2_1}
\end{align}
Eq.~\eqref{proof2_1} corresponds to Eq.~\eqref{C-MMGA}. \qed

\subsection{Proof of Theorem 3} \label{AS01-03}
We assume infinitesimal $\rd x^{a|s_i}$, and the infinitesimal change in ${\bf x}$;
\begin{align}
    &x^{a|s_i}\gets x^{a|s_i}+\rd x^{a|s_i},\\
    &{\bf x}\gets \norm({\bf x}),
\end{align}
By this change, the Markov transition matrix $\bs{M}({\bf x},{\bf y})$ changes into $\bs{M}({\bf x},{\bf y})+\rd \bs{M}({\bf x},{\bf y},\rd x^{a|s})$, described as
\begin{align}
    \rd M_{i'i''}=\rd x^{a|s_i}y^{b|s_i}\times\left\{\begin{array}{ll}
        1-x^{a|s_i} & (s_{i''}=s_i,\ s_{i'}=abs_i^{-}) \\
        -x^{a'|s_i} & (s_{i''}=s_i,\ s_{i'}=a'bs_i^{-}, a'\neq a) \\
        0 & ({\rm otherwise})
    \end{array}\right..
\end{align}
for all $i', i''\in \{1,\ldots,|\mc{S}|\}$. Then, the equilibrium state $\bs{p}^{\eq}({\bf x},{\bf y})$ changes into $\bs{p}^{\eq}({\bf x},{\bf y})+\rd \bs{p}^{\eq}({\bf x},{\bf y},\rd x^{a|s})$. Here, note that $\rd \bs{M}$ and $\rd \bs{p}^{\eq}$ are a matrix and a vector of order $O(\rd x^{a|s})$, respectively. From the stationary state condition, both $(\bs{E}-\bs{M})\bs{p}^{\eq}=0$ and $(\bs{E}-(\bs{M}+\rd\bs{M}))(\bs{p}^{\eq}+\rd\bs{p}^{\eq})=0$ hold.
\begin{align}
    &(\bs{E}-(\bs{M}+\rd\bs{M}))(\bs{p}^{\eq}+\rd\bs{p}^{\eq})-(\bs{E}-\bs{M})\bs{p}^{\eq}=0\\
    &\Leftrightarrow (\bs{E}-\bs{M})\rd\bs{p}^{\eq}=\rd\bs{M}\bs{p}^{\eq}+\bs{1}\times O((\rd x^{a|s})^2).
\end{align}
Here, the term of $O((\rd x^{a|s})^2)$ is small enough to be ignored. Then, the rest term $\delta\bs{M}\bs{p}^{\eq}$ is calculated as
\begin{align}
    (\rd\bs{M}\bs{p}^{\eq})_{i'}&=\sum_{i''}\rd M_{i'i''}p_{i''}^{\eq}=\rd M_{i'i}p_{i}^{\eq}\\
    &=\rd x^{a|s_i}p_{i}^{\eq}y^{b|s_i}\times\left\{\begin{array}{ll}
        1-x^{a|s_i} & (s_{i'}=abs_i^{-}) \\
        -x^{a'|s_i} & (s_{i'}=a'bs_i^{-}, a'\neq a) \\
        0 & ({\rm otherwise})
    \end{array}\right..\\
    &=\rd x^{a|s_i}p_{i}^{\eq}\left(p_{i'}^{a|s_i}-\sum_{a'}x^{a'|s_i}p_{i'}^{a'|s_i}\right).
\end{align}
Thus,
\begin{align}
    &(\bs{E}-\bs{M})\rd\bs{p}^{\eq}=\rd x^{a|s_i}p_{i}^{\eq}\left(\bs{p}^{a|s_i}-\sum_{a'}x^{a'|s_i}\bs{p}^{a'|s_i}\right)\\
    &\rd\bs{p}^{\eq}=\rd x^{a|s_i}p_{i}^{\eq}(\bs{E}-\bs{M})^{-1}\left(\bs{p}^{a|s_i}-\sum_{a'}x^{a'|s_i}\bs{p}^{a'|s_i}\right)\\
    &\Leftrightarrow \frac{\rd\bs{p}^{\eq}}{\rd x^{a|s_i}}=\frac{\partial}{\partial x^{a|s_i}}\bs{p}^{\eq}(\norm({\bf x}),{\bf y})=p_{s_i}^{\eq}(\bs{E}-\bs{M})^{-1}\left(\bs{p}^{a|s_i}-\sum_{a'}x^{a'|s_i}\bs{p}^{a'|s_i}\right)\\
    &\Leftrightarrow \frac{\partial}{\partial x^{a|s_i}}\bs{p}^{\eq}(\norm({\bf x}),{\bf y})\cdot\bs{u}=p_{i}^{\eq}(\bs{E}-\bs{M})^{-1}\left(\bs{p}^{a|s_i}-\sum_{a'}x^{a'|s_i}\bs{p}^{a'|s_i}\right)\cdot\bs{u}\\
    &\Leftrightarrow \frac{\partial}{\partial x^{a|s_i}}u^{\eq}(\norm({\bf x}),{\bf y})=p_{i}^{\eq}\left(\pi(\bs{p}^{a|s_i},{\bf x},{\bf y})-\sum_{a'}x^{a'|s_i}\pi(\bs{p}^{a'|s_i},{\bf x},{\bf y})\right).
    \label{proof3_13}
\end{align}
The left-hand (resp. right-hand) side of Eq.~\eqref{proof3_13} corresponds to continualized MMGA (resp. MMRD). \qed

\subsection{Proof of Theorem 4} \label{AS01-04}
Let us prove that X's strategy in Nash equilibrium is uniquely $\bs{x}=x^*\bs{1}$. First, we define $u^*$ and $v^*$;
\begin{align}
    u^*&=x^*y^*u_1+x^*(1-y^*)u_2+(1-x^*)y^*u_3+(1-x^*)(1-y^*)u_4 \\
    &=\frac{u_1u_4-u_2u_3}{u_1-u_2-u_3+u_4}\\
    (&=-v^*)
\end{align}
as X's and Y's payoffs in the Nash equilibrium in the zero-memory game. If X uses the Nash equilibrium strategy $\bs{x}=x^*\bs{1}$, the stationary state condition $\bs{p}^{\eq}=\bs{M}\bs{p}^{\eq}$ satisfies
\begin{align}
    &\bs{p}^{\eq}=\left(\begin{array}{cccc}
        x^*y_1 & x^*y_2 & x^*y_3 & x^*y_4 \\
        x^*\tilde{y}_{1} & x^*\tilde{y}_{2} & x^*\tilde{y}_{3} & x^*\tilde{y}_{4} \\
        \tilde{x}^*y_1 & \tilde{x}^*y_2 & \tilde{x}^*y_3 & \tilde{x}^*y_4 \\
        \tilde{x}^*\tilde{y}_{1} & \tilde{x}^*\tilde{y}_{2} & \tilde{x}^*\tilde{y}_{3} & \tilde{x}^*\tilde{y}_{4} \\
    \end{array}\right)\bs{p}^{\eq}\\
    &\Rightarrow \bs{p}^{\eq}=(x^*\bs{p}^{\eq}\cdot\bs{y},x^*(1-\bs{p}^{\eq}\cdot\bs{y}),\tilde{x}^*\bs{p}^{\eq}\cdot\bs{y},\tilde{x}^*(1-\bs{p}^{\eq}\cdot\bs{y}))^{\mathrm{T}} \\
    &\Rightarrow \bs{p}^{\eq}\cdot\bs{u}=u^{*} \\
    &\Leftrightarrow u^{\eq}=u^{*}.
\end{align}
which shows that X's payoff in the stationary state is $u^{*}$, regardless of Y's strategy $\bs{y}$.

Below, we show that if X uses another strategy $\bs{x}\neq x^*\bs{1}$, there always is Y's strategy such that $v^{\eq}>v^*\Leftrightarrow u^{\eq}<u^*$. As X's non-equilibrium strategy, we assume the case $x_1\neq x^*$ representatively. Then, Y's strategy $\bs{y}=y^*\bs{1}+\rd y_1\bs{e}_1$ with sufficiently small $\rd y_1$ satisfies
\begin{align}
    &\bs{p}^{\eq}=\left(\begin{array}{cccc}
        x_1(y^*+\rd y_1) & x_2y^* & x_3y^* & x_4y^* \\
        x_1(\tilde{y}^*-\rd y_1) & x_2\tilde{y}^* & x_3\tilde{y}^* & x_4\tilde{y}^* \\
        \tilde{x}_1(y^*+\rd y_1) & \tilde{x}_2y^* & \tilde{x}_3y^* & \tilde{x}_4y^* \\
        \tilde{x}_1(\tilde{y}^*-\rd y_1) & \tilde{x}_2\tilde{y}^* & \tilde{x}_3\tilde{y}^* & \tilde{x}_4\tilde{y}^* \\
    \end{array}\right)\bs{p}^{\eq}.
\end{align}
In this equation, we approximate $\bs{p}^{\eq}\simeq \bs{p}^{\eq(0)}+\bs{p}^{\eq(1)}$, where $\bs{p}^{\eq(k)}$ describes the $O((\rd y_1)^k)$ term in $\bs{p}^{\eq}$. We can derive these $0$-th and $1$-st order terms by comparing the left-hand and right-side of this equation. Here, the $0$-th order term satisfies $\bs{p}^{\eq(0)}\cdot\bs{u}=u^*$, which means that the term does not contribute to the deviation from the Nash equilibrium payoff. On the other hand, the $1$-st order term gives
\begin{align}
    &\bs{p}^{\eq(1)}=p_1^{\eq(0)}\rd y_1(+x_1,-x_1,+\tilde{x}_1,-\tilde{x}_1)^{\mathrm{T}}\\
    &\Rightarrow v^{\eq(1)}=p_1^{\eq(0)}\rd y_1\underbrace{(v_1-v_2-v_3+v_4)}_{=\bs{v}\cdot\bs{1}_z\neq 0}(x_1-x^*).
\end{align}
Here, we use $\bs{1}_z:=(+1,-1,-1,+1)$. Thus, in the leading order, $v^{\eq(1)}>v^*\Leftrightarrow u^{\eq(1)}<u^*$ holds by taking $\rd y_1>0$ if $\bs{v}\cdot\bs{1}_z(x_1-x^*)>0$, while by taking $\rd y_1<0$ if $\bs{v}\cdot\bs{1}_z(x_1-x^*)<0$. In other words, X's minimax strategy is $\bs{x}=x^*\bs{1}$. Similarly, we can prove that Y's minimax strategy is $\bs{y}=y^*\bs{1}$. Thus, the Nash equilibrium is given by $(\bs{x},\bs{y})=(x^*\bs{1},y^*\bs{1})$. \qed

\section{Analysis of Learning Dynamics} \label{AS02}
\subsection{Simpler MMGA for Two-action Games} \label{AS02-01}
This section is concerned with the contents in {\bf Section~\ref{S04-02}} in the main manuscript.

Especially in two-action games, we can use the formulation of Assumption~1 in the main manuscript. By replacing the strategies $({\bf x},{\bf y})$ by $(\bs{x},\bs{y})$, we can formulation another simpler algorithm of MMGA as
\begin{algorithm}[H]
    \caption{Discretized MMGA for two-action}
    \label{D-MMGA-TWOACTION}
    \textbf{Input}: $\eta$, $\gamma$
    \begin{algorithmic}[1]
        \FOR{$t=0,1,2,\cdots$}
        \label{AA1-L01}
        \FOR{$i=1,2,\ldots,|\mc{S}|$}
        \label{AA1-L02}
        \STATE $\bs{x}'\gets \bs{x}+\gamma\bs{e}_i$
        \label{AA1-L03}
        \STATE $\displaystyle \Delta_i\gets(1-x_i)\frac{u^{\eq}(\bs{x}',\bs{y})-u^{\eq}(\bs{x},\bs{y})}{\gamma}$
        \label{AA1-L04}
        \ENDFOR
        \label{AA1-L05}
        \FOR{$i=1,2,\ldots,|\mc{S}|$}
        \label{AA1-L06}
        \STATE $x_i\gets x_i(1+\eta\Delta_i)$
        \label{AA1-L07}
        \ENDFOR
        \label{AA1-L08}
        \ENDFOR
        \label{AA1-L09}
    \end{algorithmic}
\end{algorithm}
There is a major difference between the original and simpler MMGAs in lines~\ref{AA1-L03} and~\ref{AA1-L04}. The difference between these algorithms is resolved as
\begin{align}
    &\left\{\begin{array}{l}
        x'^{a_1|s_i}\gets x'^{a_1|s_i}+\gamma \\
        {\bf x}'\gets \norm({\bf x}') \\
        \displaystyle \Delta^{a_1|s_i}\gets\frac{u^{\eq}({\bf x}',{\bf y})-u^{\eq}({\bf x},{\bf y})}{\gamma} \\
    \end{array}\right.
    \label{proof_smpMMGA1}\\
    &\Leftrightarrow \left\{\begin{array}{l}
        x'^{a_1|s_i}\gets x'^{a_1|s_i}+(1-x'^{a_1|s_i})\gamma+O(\gamma^2) \\
        x'^{a_2|s_i}\gets x'^{a_2|s_i}-x'^{a_2|s_i}\gamma+O(\gamma^2) \\
        \displaystyle \Delta^{a_1|s_i}\gets\frac{u^{\eq}({\bf x}',{\bf y})-u^{\eq}({\bf x},{\bf y})}{\gamma} \\
    \end{array}\right.
    \label{proof_smpMMGA2}\\
    &\Leftrightarrow \left\{\begin{array}{l}
        x'_i\gets x'_i+(1-x'_i)\gamma \\
        \displaystyle \Delta_i\gets\frac{u^{\eq}(\bs{x}',\bs{y})-u^{\eq}(\bs{x},\bs{y})}{\gamma} \\
    \end{array}\right.
    \label{proof_smpMMGA3}\\
    &\Leftrightarrow \left\{\begin{array}{l}
        x'_i\gets x'_i+\gamma \\
        \displaystyle \Delta_i\gets(1-x'_i)\frac{u^{\eq}(\bs{x}',\bs{y})-u^{\eq}(\bs{x},\bs{y})}{\gamma} \\
    \end{array}\right..
    \label{proof_smpMMGA4}
\end{align}
Here, we ignored terms of $O(\gamma^2)$ and use the definition of $\bs{x}$ (i.e., $x'_i=x'^{a_1|s_i}=1-x'^{a_2|s_i}$) between Eqs.~\eqref{proof_smpMMGA2} and \eqref{proof_smpMMGA3}. Thus, we use the continualized version of this algorithm;
\begin{align}
    \dot{\bs{x}}=\bs{x}\circ(\bs{1}-\bs{x})\circ\frac{\partial}{\partial \bs{x}}u^{\eq}(\bs{x},\bs{y}).
    \label{TA-C-MMGA}
\end{align}

\subsection{Approximation of learning dynamics} \label{AS02-02}
In {\bf Section~\ref{S04-02}} and~{\bf \ref{S05-01}}, we introduce a method to approximate the learning dynamics up to $k$-th order terms for deviations from the Nash equilibrium. The stationary state condition of the one-memory two-action game is given by
\begin{align}
    &\bs{p}^{\eq}=\bs{M}\bs{p}^{\eq},\\
    &\bs{M}=\left(\begin{array}{cccc}
        x_1y_1 & x_2y_2 & x_3y_3 & x_4y_4 \\
        x_1\tilde{y}_1 & x_2\tilde{y}_2 & x_3\tilde{y}_3 & x_4\tilde{y}_4 \\
        \tilde{x}_1y_1 & \tilde{x}_2y_2 & \tilde{x}_3y_3 & \tilde{x}_4y_4 \\
        \tilde{x}_1\tilde{y}_1 & \tilde{x}_2\tilde{y}_2 & \tilde{x}_3\tilde{y}_3 & \tilde{x}_4\tilde{y}_4 \\
    \end{array}\right).
\end{align}
Here, for any variable $\mc{X}$, we define $\tilde{\mc{X}}:=1-\mc{X}$. In addition, let us denote $O(\de^k)$ term in any variable $\mc{X}$ as $\mc{X}^{(k)}$. The neighbor of the Nash equilibrium, by substituting $\bs{x}=x^*\bs{1}+\bs{\de}$ and $\bs{y}=y^*\bs{1}+\bs{\ep}$, we can decompose $\bs{M}=\sum_{k=0}^2 \bs{M}^{(k)}$ as
\begin{align}
    &\bs{M}=\underbrace{(\bs{x}^*\circ\bs{y}^*)\otimes\bs{1}}_{=\bs{M}^{(0)}}+\underbrace{(\bs{y}^*\circ\bs{1}_x)\otimes\bs{\de}+(\bs{x}^*\circ\bs{1}_y)\otimes\bs{\ep}}_{=\bs{M}^{(1)}}+\underbrace{\bs{1}_z\otimes(\bs{\de}\circ\bs{\ep})}_{=\bs{M}^{(2)}},\\
    &\bs{x}^{*}:=(x^*,x^*,\tilde{x}^*,\tilde{x}^*),\hspace{0.5cm} \bs{y}^{*}:=(y^*,\tilde{y}^*,y^*,\tilde{y}^*),\\
    &\bs{1}_x:=(+1,+1,-1,-1)^{\rm{T}},\hspace{0.5cm} \bs{1}_y:=(+1,-1,+1,-1)^{\rm{T}},\hspace{0.5cm} \bs{1}_z:=\bs{1}_x\circ\bs{1}_y=(+1,-1,-1,+1)^{\rm{T}}.
\end{align}
In the same way, we can decompose $\bs{p}^{\eq}\simeq\sum_{k=0}\bs{p}^{\eq(k)}$ as
\begin{align}
    \bs{p}^{\eq(0)}&=\bs{M}^{(0)}\bs{p}^{\eq(0)}=\underbrace{(\bs{p}^{\eq(0)}\cdot\bs{1})}_{=1}\bs{x}^*\circ\bs{y}^*=:\bs{p}^*,\\
    \bs{p}^{\eq(1)}&=\bs{M}^{(1)}\bs{p}^{\eq(0)}=(\bs{\de}\cdot \bs{p}^*)\bs{y}^*\circ\bs{1}_x+(\bs{\ep}\cdot \bs{p}^*)\bs{x}^*\circ\bs{1}_y,\\
    \bs{p}^{\eq(2)}&=\bs{M}^{(2)}\bs{p}^{\eq(0)}+\bs{M}^{(1)}\bs{p}^{\eq(1)}\\
    &=(\bs{\de}\circ\bs{\ep}\cdot\bs{p}^*)\bs{1}_z+\left\{(\bs{\de}\cdot\bs{p}^*)(\bs{\de}\circ\bs{y}^*\cdot\bs{1}_x)+(\bs{\ep}\cdot\bs{p}^*)(\bs{\de}\circ\bs{x}^*\cdot\bs{1}_y)\right\}\bs{y}^*\circ\bs{1}_x
    \nonumber\\
    &\hspace{0.4cm}+\left\{(\bs{\de}\cdot\bs{p}^*)(\bs{\ep}\circ\bs{y}^*\cdot\bs{1}_x)+(\bs{\ep}\cdot\bs{p}^*)(\bs{\ep}\circ\bs{x}^*\cdot\bs{1}_y)\right\}\bs{x}^*\circ\bs{1}_y.
\end{align}
We also get $\bs{p}^{\eq(k)}=\bs{M}^{(2)}\bs{p}^{\eq(k-2)}+\bs{M}^{(1)}\bs{p}^{\eq(k-1)}$ for further orders of $k\ge 2$. Then, the equilibrium payoff is given by
\begin{align}
    u^{\eq(0)}&=\bs{p}^{\eq(0)}\cdot\bs{u}=u^{\eq},\\
    u^{\eq(1)}&=\bs{p}^{\eq(1)}\cdot\bs{u}=0,\\
    u^{\eq(2)}&=\bs{p}^{\eq(2)}\cdot\bs{u}=(\bs{u}\cdot\bs{1}_z)(\bs{\de}\circ\bs{\ep}\cdot\bs{p}^*),\\
    u^{\eq(3)}&=\bs{p}^{\eq(3)}\cdot\bs{u}=(\bs{M}^{(2)}\bs{p}^{\eq(1)}+\bs{M}^{(1)}\bs{p}^{\eq(2)})\cdot\bs{u}=(\bs{u}\cdot\bs{1}_z)(\bs{\de}\circ\bs{\ep}\cdot\bs{p}^{\eq(1)})\\
    &=(\bs{u}\cdot\bs{1}_z)\left\{(\bs{\de}\cdot\bs{p}^*)(\bs{\de}\circ\bs{\ep}\circ\bs{y}^*\cdot\bs{1}_x)+(\bs{\ep}\cdot\bs{p}^*)(\bs{\de}\circ\bs{\ep}\circ\bs{x}^*\cdot\bs{1}_y)\right\},\\
    u^{\eq(4)}&=\bs{p}^{\eq(4)}\cdot\bs{u}=(\bs{M}^{(2)}\bs{p}^{\eq(2)}+\bs{M}^{(1)}\bs{p}^{\eq(3)})\cdot\bs{u}=(\bs{u}\cdot\bs{1}_z)(\bs{\de}\circ\bs{\ep}\cdot\bs{p}^{\eq(2)})\\
    &=(\bs{u}\cdot\bs{1}_z)[(\bs{\de}\circ\bs{\ep}\cdot\bs{p}^*)(\bs{\de}\circ\bs{\ep}\cdot\bs{1}_z)
    \nonumber\\
    &\hspace{1.7cm}+\{(\bs{\de}\cdot\bs{p}^*)(\bs{\de}\circ\bs{y}^*\cdot\bs{1}_x)+(\bs{\ep}\cdot\bs{p}^*)(\bs{\de}\circ\bs{x}^*\cdot\bs{1}_y)\}(\bs{\de}\circ\bs{\ep}\circ\bs{y}^*\cdot\bs{1}_x)
    \nonumber\\
    &\hspace{1.7cm}+\{(\bs{\de}\cdot\bs{p}^*)(\bs{\ep}\circ\bs{y}^*\cdot\bs{1}_x)+(\bs{\ep}\cdot\bs{p}^*)(\bs{\ep}\circ\bs{x}^*\cdot\bs{1}_y)\}(\bs{\de}\circ\bs{\ep}\circ\bs{x}^*\cdot\bs{1}_y)].
\end{align}
Here, we used
\begin{align}
    \bs{M}^{(1)\mathrm{T}}\bs{u}&=\underbrace{(\bs{y}^*\circ\bs{1}_x\cdot\bs{u})}_{=0}\bs{\de}+\underbrace{(\bs{x}^*\circ\bs{1}_y\cdot\bs{u})}_{=0}\bs{\ep}=\bs{0},\\
    \bs{M}^{(2)\mathrm{T}}\bs{u}&=(\bs{u}\cdot\bs{1}_z)\bs{\de}\circ\bs{\ep}.
\end{align}
Then, the gradient of this payoff is given by
\begin{align}
    \frac{\partial u^{\eq(1)}}{\partial \bs{\de}}&=\bs{0},\\
    \frac{\partial u^{\eq(2)}}{\partial \bs{\de}}&=(\bs{u}\cdot\bs{1}_z)\bs{\ep}\circ\bs{p}^*,\\
    \frac{\partial u^{\eq(3)}}{\partial \bs{\de}}&=(\bs{u}\cdot\bs{1}_z)\{(\bs{\de}\cdot\bs{p}^*)\bs{\ep}\circ\bs{y}^*\circ\bs{1}_x+(\bs{\ep}\cdot\bs{p}^*)\bs{\ep}\circ\bs{x}^*\circ\bs{1}_y+(\bs{\de}\circ\bs{\ep}\circ\bs{y}^*\cdot\bs{1}_x)\bs{p}^*\},\\
    \frac{\partial u^{\eq(4)}}{\partial \bs{\de}}&=(\bs{u}\cdot\bs{1}_z)[(\bs{\de}\circ\bs{\ep}\cdot\bs{p}^*)(\bs{\ep}\circ\bs{1}_z)+(\bs{\de}\circ\bs{\ep}\cdot\bs{1}_z)(\bs{\ep}\circ\bs{p}^*)
    \nonumber\\
    &\hspace{1.7cm}+\{(\bs{\de}\cdot\bs{p}^*)(\bs{\de}\circ\bs{y}^*\cdot\bs{1}_x)+(\bs{\ep}\cdot\bs{p}^*)(\bs{\de}\circ\bs{x}^*\cdot\bs{1}_y)\}(\bs{\ep}\circ\bs{y}^*\circ\bs{1}_x)
    \nonumber\\
    &\hspace{1.7cm}+\{(\bs{\de}\cdot\bs{p}^*)(\bs{\ep}\circ\bs{y}^*\cdot\bs{1}_x)+(\bs{\ep}\cdot\bs{p}^*)(\bs{\ep}\circ\bs{x}^*\cdot\bs{1}_y)\}(\bs{\ep}\circ\bs{x}^*\circ\bs{1}_y)
    \nonumber\\
    &\hspace{1.7cm}+(\bs{\de}\circ\bs{\ep}\circ\bs{y}^*\cdot\bs{1}_x)\{(\bs{\de}\cdot\bs{p}^*)\bs{y}^*\circ\bs{1}_x+(\bs{\ep}\cdot\bs{p}^*)\bs{x}^*\circ\bs{1}_y+(\bs{\de}\circ\bs{y}^*\cdot\bs{1}_x)\bs{p}^*\}
    \nonumber\\
    &\hspace{1.7cm}+(\bs{\de}\circ\bs{\ep}\circ\bs{x}^*\cdot\bs{1}_y)(\bs{\ep}\circ\bs{y}^*\cdot\bs{1}_x)\bs{p}^*].
\end{align}

The learning dynamics (of continualized MMGA) in two-action one-memory games are given by
\begin{align}
    \dot{\bs{\de}}&=+\left(x^*\bs{1}+\bs{\de}\right)\circ\left(\tilde{x}^*\bs{1}-\bs{\de}\right)\circ\frac{\partial u^{\eq}}{\partial\bs{\de}},\\
    \dot{\bs{\ep}}&=-\left(y^*\bs{1}+\bs{\ep}\right)\circ\left(\tilde{y}^*\bs{1}-\bs{\ep}\right)\circ\frac{\partial u^{\eq}}{\partial\bs{\ep}}.
\end{align}
We can decompose $\dot{\bs{\de}}\simeq\sum_{k=0}\dot{\bs{\de}}^{(k)}$ and $\dot{\bs{\ep}}\simeq\sum_{k=0}\dot{\bs{\ep}}^{(k)}$ as
\begin{align}
    \dot{\bs{\de}}^{(0)}&=+x^*\tilde{x}^*\frac{\partial u^{\eq(1)}}{\partial\bs{\de}},\\
    \dot{\bs{\de}}^{(1)}&=+x^*\tilde{x}^*\frac{\partial u^{\eq(2)}}{\partial\bs{\de}}-(x^*-\tilde{x}^*)\bs{\de}\circ\frac{\partial u^{\eq(1)}}{\partial\bs{\de}},\\
    \dot{\bs{\de}}^{(2)}&=+x^*\tilde{x}^*\frac{\partial u^{\eq(3)}}{\partial\bs{\de}}-(x^*-\tilde{x}^*)\bs{\de}\circ\frac{\partial u^{\eq(2)}}{\partial\bs{\de}}+\bs{\de}\circ\bs{\de}\circ\frac{\partial u^{\eq(1)}}{\partial\bs{\de}},\\
    \dot{\bs{\de}}^{(3)}&=+x^*\tilde{x}^*\frac{\partial u^{\eq(4)}}{\partial\bs{\de}}-(x^*-\tilde{x}^*)\bs{\de}\circ\frac{\partial u^{\eq(3)}}{\partial\bs{\de}}+\bs{\de}\circ\bs{\de}\circ\frac{\partial u^{\eq(2)}}{\partial\bs{\de}},\\
    \dot{\bs{\ep}}^{(0)}&=-y^*\tilde{y}^*\frac{\partial u^{\eq(1)}}{\partial\bs{\ep}},\\
    \dot{\bs{\ep}}^{(1)}&=-y^*\tilde{y}^*\frac{\partial u^{\eq(2)}}{\partial\bs{\ep}}+(y^*-\tilde{y}^*)\bs{\ep}\circ\frac{\partial u^{\eq(1)}}{\partial\bs{\ep}},\\
    \dot{\bs{\ep}}^{(2)}&=-y^*\tilde{y}^*\frac{\partial u^{\eq(3)}}{\partial\bs{\ep}}+(y^*-\tilde{y}^*)\bs{\ep}\circ\frac{\partial u^{\eq(2)}}{\partial\bs{\ep}}-\bs{\ep}\circ\bs{\ep}\circ\frac{\partial u^{\eq(1)}}{\partial\bs{\ep}},\\
    \dot{\bs{\ep}}^{(3)}&=-y^*\tilde{y}^*\frac{\partial u^{\eq(4)}}{\partial\bs{\ep}}+(y^*-\tilde{y}^*)\bs{\ep}\circ\frac{\partial u^{\eq(3)}}{\partial\bs{\ep}}-\bs{\ep}\circ\bs{\ep}\circ\frac{\partial u^{\eq(2)}}{\partial\bs{\ep}}.
\end{align}
In cases of one-memory penny-matching games, the solution is obtained if we substitute
\begin{align}
    \bs{x}^*=\frac{1}{2}\bs{1},\hspace{0.5cm} \bs{y}^*=\frac{1}{2}\bs{1}, \hspace{0.5cm} \bs{p}^{*}=\frac{1}{4}\bs{1}.
\end{align}

\subsection{Method to Calculate the Stationary State} \label{AS02-03}
Regarding {\bf Section~\ref{S05-01}}, we use an analytical solution of the stationary state, which is known only in the case of two-action one-memory games as
\begin{align}
    &p_1^{\eq}=k\{(x_{4}+(x_{3}-x_{4})y_{3})(y_{4}+(y_{2}-y_{4})x_{2})-x_{3}y_{2}(x_{2}-x_{4})(y_{3}-y_{4})\}, \\
	&p_2^{\eq}=k\{(x_{4}+(x_{3}-x_{4})y_{4})(\tilde{y}_{3}-(y_{1}-y_{3})x_{1})-x_{4}\tilde{y}_{1}(x_{1}-x_{3})(y_{3}-y_{4})\}, \\
	&p_3^{\eq}=k\{(\tilde{x}_{2}-(x_{1}-x_{2})y_{1})(y_{4}+(y_{2}-y_{4})x_{4})-\tilde{x}_{1}y_{4}(x_{2}-x_{4})(y_{1}-y_{2})\}, \\
	&p_4^{\eq}=k\{(\tilde{x}_{2}-(x_{1}-x_{2})y_{2})(\tilde{y}_{3}-(y_{1}-y_{3})x_{3})-\tilde{x}_{2}\tilde{y}_{3}(x_{1}-x_{3})(y_{1}-y_{2})\}, \\
    &k=\frac{1}{p_1^{\eq}+p_2^{\eq}+p_3^{\eq}+p_4^{\eq}},
\end{align}
under the notation in Assumption~\ref{Assumption01}.

In other parts ({\bf Sections~\ref{S05-02}} and~{\bf \ref{S05-03}}), we calculate the stationary state of a Markov transition matrix $\bs{M}$ by the power iteration method. Compared to the analytical solution $\bs{p}^{\eq}$, the computational solution $\hat{\bs{p}}^{\eq}$ is accurate except for $10^{-9}$ error in $L^2$ norm (i.e., $\|\hat{\bs{p}}^{\eq}-\bs{p}^{\eq}\|_2\le 10^{-9}$).

\section{Divergence from Nash Equilibrium Unaffected by Numerical Errors} \label{AS03}
This section is devoted to proving that a finding of this study, i.e., the divergence from the Nash equilibrium, is not due to the accumulation of errors in our finite difference method. When we implemented continualized MMGA, we used the fourth-order Runge-Kutta method with the step size of $10^{-2}$. In this method, the accumulation of simulation errors can be estimated as sufficiently small ($O(10^{-2})^{4}=O(10^{-8})$ per unit of time). 

\begin{figure}[ht]
    \centering
    \includegraphics[width=0.8\hsize]{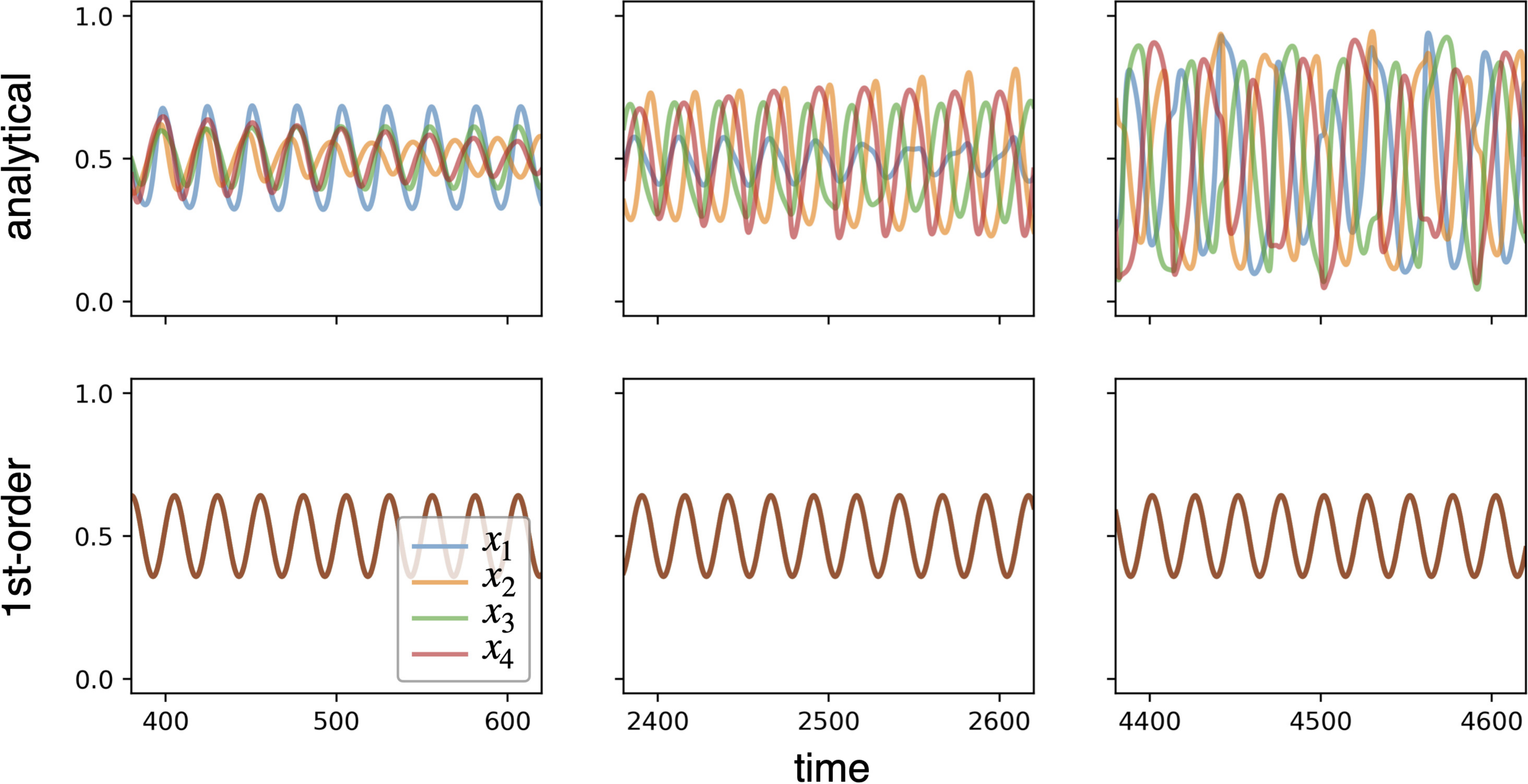}
    \caption{The analytical dynamics of continualized MMGA (the upper panels) and the first-order approximation (the lower panels). Initial conditions are the same between the upper and lower panels ($\bs{x}=\bs{y}=0.6\times \bs{1}$).}
    \label{FS01}
\end{figure}

We also numerically demonstrate that divergence is true. Fig.~\ref{FS01} compares the analytical dynamics of continualized MMGA (Eq.~\eqref{C-MMGA} in the main manuscript) and the first-order approximation of this continualized MMGA. Both the dynamics are output by the fourth-order Runge-Kutta method with the step size of $10^{-2}$. While the analytical dynamics gradually diverge from the Nash equilibrium, the first-order approximation continues to draw circular orbits after a sufficiently long time has passed. Thus, the figure shows that this divergence is due to higher-order terms in continualized MMGA.

\end{document}